\documentclass[final,1p,times]{elsarticle}

\usepackage{graphicx}
\usepackage{amssymb}
\usepackage{amsthm}
\usepackage{lineno}

\journal{Nuclear Physics A}

\begin{document}

\begin{frontmatter}

\title{Characterizing the hydrodynamic response to the initial conditions}


\author[auth1]{Fernando G.  Gardim}
\author[auth1]{Fr\'ed\'erique Grassi}
\author[auth2,mcgill,lbnl]{Matthew Luzum}
\author[auth2]{Jean-Yves Ollitrault}
\address[auth1]{Instituto de F\'\i sica, Universidade de S\~ao Paulo, C.P. 66318, 05315-970, S\~ao Paulo-SP, Brazil}
\address[auth2]{CNRS, URA2306, IPhT, Institut de physique th\'eorique de Saclay, F-91191
Gif-sur-Yvette, France}
\address[mcgill]{McGill University,  3600 University Street, Montreal QC H3A 2TS, Canada}
\address[lbnl]{Lawrence Berkeley National Laboratory, Berkeley, CA 94720, USA}
\begin{abstract}

In hydrodynamics, the momentum distribution of particles at the end of the evolution is completely determined by initial conditions. We study quantitatively to what extent anisotropic flow $v_n$ is determined by predictors such as the initial eccentricity $\varepsilon_n$ in a set of realistic simulations, and we also show the importance of nonlinear terms in order to correctly predict $v_4$. This knowledge will be important for making a more direct link between experimental observables and hydrodynamic initial conditions, the latter being poorly constrained at present.  

\end{abstract}

\end{frontmatter} 


\section{Introduction}

One of the most important probes of ultrarelativistic heavy-ion collisions is the anisotropic flow. Event-by-event hydrodynamics \cite{Hama:2004rr} provides a natural way of studying anisotropic flow and its fluctuations.  In such calculations, one supplies a set of initial conditions (IC), evolves them through ideal or viscous hydrodynamics, and finally computes particle emission at the end. One can then write the azimuthal distribution of outgoing particles in a hydrodynamic event as a Fourier series
\begin{eqnarray}
\frac{2\pi}{N}\frac{dN}{d\phi}=1+2\sum_n^\infty v_n\cos[n(\phi-\Psi_n)],
\label{eq:v-def}
\end{eqnarray}
or equivalently $\left\{e^{i n\phi}\right\}=v_ne^{-i n\Psi_n}$, where $\{\cdots\}$ is the average in one event. Since the largest source of uncertainty in these calculations is the initial conditions,
it is useful to identify which properties of the initial state determine each observable.  This knowledge can then allow to constrain the initial state directly from data.

It is well known that the initial average profile of non-central collisions is almond-shaped. One can obtain the eccentricity and the direction of the initial profile computing $\varepsilon_2e^{i2\Phi_2}=-\left\{r^2e^{i2\phi}\right\}/\left\{r^2\right\}$ in the CM frame, where $\varepsilon_2$ is the participant eccentricity, $\Phi_2$ is the participant plane, whose direction is  the minor axis of the ellipse, and $\{\cdots\}$ denotes an average over the initial density profile. The hydrodynamic expansion produces an elliptic distribution of particles, i.e. $\varepsilon_2\propto v_2$, and the  maximum of the distribution of particles is aligned with the direction of the steepest energy gradient in the IC, i.e $\Phi_2=\Psi_2$. Nevertheless, due to the finite number of nucleons colliding in each nucleus-nucleus collisions, there are fluctuations in the initial profile, thus the relations $v_2\propto\varepsilon_2$ and $\Psi_2=\Phi_2$ may no longer be valid in event-by-event hydrodynamics. However, the study of these relations for several models of IC in event-by-event hydrodynamics, showed that they are reasonably satisfied \cite{Gardim:2011re,Petersen:2010cw,Qin:2010pf,Qiu:2011iv}.


Symmetry considerations have been used to argue that higher-order flows ($v_n$ with $n\ge 3$) should also be created by an anisotropy, $\varepsilon_n$. It has been shown that this is also true for $n=3$, but it is not valid for $n=4,5$~\cite{Qin:2010pf,Qiu:2011iv}. Nevertheless, Teaney and Yan \cite{Teaney:2010vd} have introduced a cumulant expansion in the initial density profile, in which $\varepsilon_n$ is only the first term in an infinite series, and they have suggested that the hydrodynamic response may be improved by adding higher-order or nonlinear terms. 

In order to understand which properties of the initial state determine $v_n$ and $\Psi_n$, 
we propose a simple quantitative measure of the correlation between $\left(v_n,\Psi_n\right)$ and any proposed predictor, such as $\left(\varepsilon_n,\Phi_n\right)$. We then use this to find better estimators for the anisotropic flows.  Details can be found in \cite{Gardim:2011xv}.
\section{Characterizing the hydrodynamic response}

In previous works~\cite{Petersen:2010cw,Qin:2010pf,Qiu:2011iv}, the correlation of the anisotropic flow with the initial geometry has been studied by visually inspecting two types of plots representing separately magnitude and direction: the $\Psi_n-\Phi_n$ distribution, and a scatter plot of $v_n$ versus $\varepsilon_n$. Instead, we carry out a global analysis, studying the correlation between the entire vector both simultaneously and quantitatively. For a given event in a centrality class, we write

\begin{equation}
v_n e^{in\Psi_n}=k\varepsilon_ne^{in\Phi_n}+\cal E,
\label{eq:estimator}
\end{equation}
where $k$ is a proportionality constant, and $\cal E$ is the difference between the calculated flow and the proposed estimator, or the error in the estimate.
 The generalized eccentricity $\varepsilon_n$, used throughout this work, is: $\varepsilon_ne^{i n\Phi_n}=-\left\{r^ne^{i n\phi}\right\}/\left\{r^n\right\}$,
%
%
which was proposed in Ref. \cite{Petersen:2010cw}, and later showed to be the lowest term in a cumulant expansion of the initial energy density \cite{Teaney:2010vd}. The best estimator is defined as the one that minimizes the mean square error $\langle|{\cal E}^2|\rangle$, where $\langle\cdots\rangle$ is the average over events in a centrality class. Thus, the best linear fit is achieved when 

\begin{equation}
k=\frac{\langle\varepsilon_n v_n\cos[n(\Psi_n-\Phi_n)]\rangle}{\langle\varepsilon_n^2\rangle}.
\label{eq:k}
\end{equation}
%
We define the quality of the estimator by 
\begin{equation}
Quality=k\frac{\sqrt{\langle\varepsilon_n^2\rangle}}{\sqrt{\langle v_n^2\rangle}}.
\label{eq:quality}
\end{equation}
The closer $Quality$ to 1, the better the response and the smaller the rms error $\langle|{\cal E}^2|\rangle$. Note that a negative $Quality$ means that $\Psi_n$ and $\Phi_n$ are anticorrelated. 


We now present results for Au-Au collisions at the top RHIC energy using the hydrodynamic code NeXSPheRIO~\cite{Hama:2004rr}. NeXSPheRIO evolves initial conditions generated by the event generator NeXus, solves the equations of relativistic ideal hydrodynamics, then emits particles at the end of hydrodynamical evolution using a Monte-Carlo generator. For this work, we  generated 150 events in each of the $10\%$ centrality classes, and we added 115 events in an extra class for events with zero impact parameter - particles from the pseudorapidity interval $|\eta|<1$ were used. Eccentricities are obtained by averaging over the initial transverse energy density profile at z=0. Our hydrodynamical calculations contain fluctuating initial flow, as well as longitudinal fluctuations, thus the final flow measured is not entirely determined by the initial transverse geometry. In this sense, these results represent something of a worst-case scenario.

\begin{figure}[h]
\begin{center}
 \includegraphics[width=0.58\textwidth]{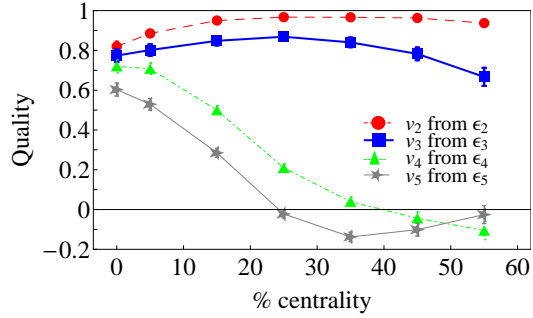}
\end{center}
\vspace{-.7cm}
\caption{\textit{Quality} of the hydrodynamic response for elliptic (circle), triangular (square), quadrangular (triangle) and pentagonal flow (star) for different centrality bins. The leftmost points correspond to events with zero impact parameter.}
\label{fig:1}
\end{figure}

Fig. \ref{fig:1} displays the \textit{Quality} of the hydrodynamic response for elliptic and triangular flow as a function of centrality. As expected, the elliptic flow is driven to the almond-shaped overlap area for non-central collisions, where $\Psi_2$ is approximately coincident with $\Phi_2$. Even for central collisions, where $v_2$ comes from fluctuations, the best estimator given by Eq. (\ref{eq:estimator}) is able to capture the physics of the elliptic flow fluctuations. The \textit{Quality} for $n=3$ is not as close to 1 as for the elliptic case, but one can see that the triangularity $\varepsilon_3$ is still a good estimator of the triangular flow.
We have also tested the triangularity definition proposed by Roland and Alver \cite{Alver:2010gr} (not shown, see Ref. \cite{Gardim:2011xv}), where $r^3$ is replaced by $r^2$ in Eq. (\ref{eq:estimator}). The triangularity with $r^3$ weight is a better predictor than with $r^2$ for most centralities. Since the $r^3$ weight gives more weight to outer shells of the initial energy profile, this means that the source of anisotropic flow moves inward as the collision becomes more peripheral.

The estimators for $n=4,5$ differ qualitatively from our results for $v_2$ and $v_3$, where  $\varepsilon_4$ and $\varepsilon_5$ only give reasonable predictions for central collisions. Moving to peripheral collisions the \textit{Quality} decreases and even become negative. 
This shows that $\varepsilon_n$ alone cannot be used to map the hydrodynamics response to the initial geometry for $n=4,5$.
\section{Finding better estimators}

One can hope to improve the estimator adding more terms in Eq. (\ref{eq:estimator}), e.g.

\begin{equation}
v_n e^{in\Psi_n}=k\varepsilon_ne^{in\Phi_n}+k'\varepsilon_n'e^{in\Phi_n'}+\cal E,
\label{eq:better}
\end{equation}
where $\varepsilon_n'$ and $\Phi_n'$ are other quantities determined from the initial density profile (for instance, the next higher cumulant). The procedure to obtain the better estimator is similar: minimize the mean-square error  $\langle|{\cal E}^2|\rangle$ with respect to $k$ and $k'$, take the average over events in each centrality class, then insert the values of $k$ and $k'$ in Eq. (\ref{eq:better}) and compute

\begin{equation}
Quality=\frac{\langle|k\varepsilon_ne^{in\Phi_n}+k'\varepsilon_n'e^{in{\Phi_n'}}|^2\rangle}{\langle{v_n^2}\rangle}.
\label{eq:nq}
\end{equation}
One can show that the quality is higher than with just one term, Eq. (\ref{eq:quality}). 

We want to apply the improved estimator to the quadrangular flow (for other flows see Ref. \cite{Gardim:2011xv}), but we need to know which are $\varepsilon_n'$ and $\Phi_n'$. Taking an elliptic initial condition, i.e an elliptic profile with only $\varepsilon_2\neq 0$, then evolving it through ideal hydrodynamics, and computing the particle emission, one will obtain, using Eq. (\ref{eq:v-def}), the anisotropic flows and the event-plane angles. Since there is only $\varepsilon_2\neq0$ one should expect only $v_2\neq0$. The results for this initial condition are: $v_2\neq0$ as a response to $\varepsilon_2$ and $\Psi_2=\Phi_2$, $v_n=0$ when $n$ odd (by symmetry), but there is a quadrangular flow, $v_4\neq0$, and $\Psi_4$ is aligned with $\Phi_2$. Since there is no $\varepsilon_4$, so where does $v_4$ come from? The answer is, $v_4$ is induced by the almond shape.

\begin{figure}[htbp]
\begin{center}
 \includegraphics[width=0.6\textwidth]{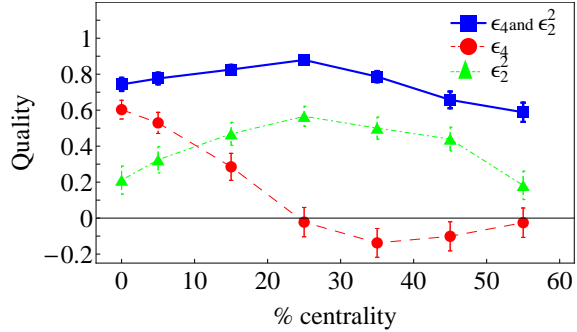}
\end{center}
\vspace{-.7cm}
\caption{\textit{Quality} for 3 different estimators of $v_4$. The combined estimator (square) is the best estimator.}
\label{fig:2}
\end{figure}

As quadrangular flow can also come from $\varepsilon_2$, we want to investigate, first, $v_4$ solely as a response to the almond shape. Thus, the natural estimator is $v_4 e^{i4\Psi_4}=k(\varepsilon_2e^{i2\Phi_2})^2+\cal E$ -- the square in the first term of rhs is used to preserve the rotational symmetry. The result can be seen in the Fig. \ref{fig:2}. For mid-central collisions, where $\varepsilon_2$ is large, the nonlinear term is important, yet this estimator is not as good as previous estimators of $v_2$ and $v_3$. Therefore, to obtain a better estimator for $v_4$ is necessary to combine both contributions, thus using Eqs. (\ref{eq:better}) and (\ref{eq:nq}), Fig. \ref{fig:2} show that the resulting estimator is good for all centralities. For $v_5$, the best estimator has linear e nonlinear terms: $\varepsilon_5$ and $\varepsilon_2\varepsilon_3$ \cite{Gardim:2011xv}.\\
%
We have defined a quantitative measure of the quality of estimators of $v_n$ from initial conditions in event-by-event hydrodynamics. $v_2$ and $v_3$ are well predicted by $\varepsilon_2$ and $\varepsilon_3$, but for $v_4$ and $v_5$ it is necessary nonlinear terms. These results provide an improved understanding of the hydrodynamic response to the initial state in realist heavy-ion collisions.

\vspace{.2cm}
\textbf{Acknowledgments:} This work is funded by FAPESP under projects 09/50180-0 and 09/16860-3, by CNPq under project 301141/2010-0, by "Agence Nationale de la Recherche" under grant ANR-08-BLAN-0093-01, and by Cofecub under project Uc Ph 113/08;2007.1.875.43.9.

\end{document}